\def\year{2020}\relax

\documentclass[letterpaper]{article} 
\usepackage{aaai20}  
\usepackage{times}  
\usepackage{comment}
\usepackage{booktabs}
\usepackage{helvet} 
\usepackage{courier}  
\usepackage[hyphens]{url}  
\usepackage{graphicx} 
\usepackage{graphicx}  
\usepackage{multirow}

\usepackage{array,booktabs,arydshln,xcolor}
\usepackage{url}

\title{On Identifying Hashtags in Disaster Twitter Data}
\author{Jishnu Ray Chowdhury$^1$, Cornelia Caragea$^1$, and Doina Caragea$^2$\\
$^1$Department of Computer Science, University of Illinois at Chicago\\
$^2$Department of Computer Science, Kansas State University\\
{\tt jraych2@uic.edu,  cornelia@uic.edu, dcaragea@ksu.edu}\\
}
\usepackage{etoolbox}
\makeatletter
\patchcmd{\maketitle}{\@copyrightspace}{}{}{}
\makeatother
 
\begin{document}
\definecolor{blue_}{HTML}{A6C5FB}
\definecolor{green_}{HTML}{00FF00}
\definecolor{orange_}{HTML}{F5DA81}
\definecolor{red_}{HTML}{F79D9B} 

\definecolor{mypink1}{rgb}{0.858, 0.188, 0.478}
\definecolor{mypink2}{RGB}{219, 48, 122}
\definecolor{mypink3}{cmyk}{0, 0.7808, 0.4429, 0.1412}
\maketitle

\begin{abstract}
Tweet hashtags have the potential to improve the search for information during disaster events. However, there is a large number of disaster-related tweets that do not have any user-provided hashtags. Moreover, only a small number of tweets that contain actionable hashtags are useful for disaster response. To facilitate progress on automatic identification (or extraction) of disaster hashtags for Twitter data, we construct a unique dataset of disaster-related tweets annotated with hashtags useful for filtering actionable information. Using this dataset, we further investigate Long Short-Term Memory-based models within a Multi-Task Learning framework. The best performing model achieves an F1-score as high as $92.22\%$. The dataset, code, and other resources are available on Github.\footnote{https://github.com/JRC1995/Tweet-Disaster-Keyphrase}
\end{abstract}

\section{Introduction}
During disasters, affected individuals often turn to social media platforms, such as Twitter and Facebook, to find the latest updates from government and response organizations, to request help or to post information that can be used to enhance situational awareness \cite{Rhodan:2017,MacMillan:2017,Frej:2018,Lapin:2018}. Nonetheless, the value of the information posted on social media platforms during disasters is highly unexploited, in part due to the lack of tools that can help filter relevant, informative, and actionable messages \cite{Villegas:2018}.

According to \citeauthor{Villegas:2018} \shortcite{Villegas:2018}, 
more than $5,200$ rescue requests made on social media were missed by the first responders, while 
about 46\% of the critical damage information posted on social media during Harvey Hurricane was missed by FEMA in their original damage estimates (that is almost half of the total costs of $\$125$ billion estimated for this hurricane).\footnote{https://coast.noaa.gov/states/fast-facts/hurricane-costs.html} As an official explained: ``It's very labor intensive to watch [social media] and because of the thousand different ways people can {\it hashtag} something or {\it keyword} something, trying to sort out what's relevant and what's not and what's actionable is very, very difficult'' \cite{SILVERMAN:2017}.

\begin{table*}[htp]
\centering
\small
\begin{tabular}{l|l}
\hline
{\bf No.}                                       & {\bf Tweet text}                                                                                             \\ \hline
\multicolumn{1}{c|}{\multirow{2}{*}{1.}} & \begin{tabular}[c]{@{}l@{}}we need help in Houston. our apartments are surrounded with water like an island we need rescue 10373 N Sam\\ Houston Pkwy E\end{tabular}                                                                                                                                \\ \cdashline{2-2} 
\multicolumn{1}{c|}{}                    & \begin{tabular}[c]{@{}l@{}} {\colorbox{red!25!white}{need help}} $|$ {\colorbox{red!25!white}{Houston}} $|$ {\colorbox{red!25!white}{need rescue}} 
\end{tabular}                  

\\ \hline \hline
\multicolumn{1}{c|}{\multirow{2}{*}{2.}} & \begin{tabular}[c]{@{}l@{}}@houstonpolice please help I'm stranded with my kids I need help fast my address is 8618 Banting st. houston \\ tx 77078.\end{tabular}                                                                                                                                   \\ \cdashline{2-2} 
\multicolumn{1}{c|}{}                    & \begin{tabular}[c]{@{}l@{}} {\colorbox{red!25!white}{stranded}} $|$ {\colorbox{red!25!white}{need help}} $|$ {\colorbox{red!25!white}{houston}}
\end{tabular}                                                                                                                                   \\ \hline \hline
\multicolumn{1}{c|}{\multirow{2}{*}{3.}}                     & \begin{tabular}[c]{@{}l@{}}Big tree fell on power lines and blocking Brown Ave near Washington St in Orlando's Thornton Park \\ neighborhood. {\color{blue}\#HurricaneIrma}\end{tabular}                                                                                                                          \\ \cdashline{2-2} 
                                          & \begin{tabular}[c]{@{}l@{}}{\colorbox{red!25!white}{power lines}} $|$ {\colorbox{red!25!white}{blocking}} $|$ {\colorbox{red!25!white}{Orlando}} $|$ {\colorbox{red!25!white}{\#HurricaneIrma}}\end{tabular}                                                                                                                          \\ \hline \hline

\multicolumn{1}{c|}{\multirow{2}{*}{4.}}                      & \begin{tabular}[c]{@{}l@{}}Very extensive damage sustained throughout {\color{blue}\#Wilmington}, {\color{blue}\#ncwx}... from {\color{blue}\#hurricane \#Florence}. Lots of trees \\ split or uprooted, siding ripped from homes, powerlines down, flooding of downtown streets, etc.\end{tabular}                                           \\ \cdashline{2-2} 
                                          & \begin{tabular}[c]{@{}l@{}} {\colorbox{red!25!white}{extensive damage}} $|$ {\colorbox{red!25!white}{\#Wilmington}} $|$ {\colorbox{red!25!white}{\#hurricane}} $|$ {\colorbox{red!25!white}{\#Florence}} $|$ {\colorbox{red!25!white}{powerlines down}} \end{tabular}                 \\ \hline \hline 
                                          
 \multicolumn{1}{c|}{\multirow{2}{*}{5.}}                     & \begin{tabular}[c]{@{}l@{}}I am evacuated from my house but I'm safe. {\color{blue}\#fire} {\color{blue}\#CampFire} {\color{blue}\#WoolseyFire \#wildfire \#safe \#Evacuation }\\ {\color{blue}  \#evacuations \#EVACUATED} {\color{blue} \#scary \#ThousandOaks \#Camarillo}\end{tabular}                                                                                   \\ \cdashline{2-2} 
                                          & \begin{tabular}[c]{@{}l@{}} {\colorbox{red!25!white}{evacuated}} $|$ 
                                    {\colorbox{red!25!white}{\#WoolseyFire}} $|$ {\colorbox{red!25!white}{\#ThousandOaks}} $|$ {\colorbox{red!25!white}{\#Camarillo}}
                                          \end{tabular}                                                                                   \\ \hline                                         
                                          
\end{tabular}
\caption{Examples of tweets posted during disasters. The original user-provided hashtags, when available, are shown in blue color for each tweet. Relevant and topically more informative hashtags manually identified to have the potential to retrieve actionable disaster tweets are highlighted in a light red box.
}
\label{examples}
\end{table*}

Examples of tweets that illustrate the diverse ways in which people use hashtags to highlight information during disasters are shown in Table \ref{examples}. Specifically, 
the user-provided hashtags, when available, are shown in blue color in the table. 
As can be seen, the first two tweets do not have any user-provided hashtags. The third tweet  has a general disaster-name hashtag,  \#HurricaneIrma. While this hashtag is useful in recognizing that the tweet was posted during Hurricane Irma, it is not useful in identifying situational awareness (e.g., damage, power loss, blocked street) or the type of disaster response requests. The fourth tweet, which explicitly reports damage, has disaster-name, location, and  weather as hashtags, but no  specific hashtag about damage. 
Finally, the fifth tweet is at the other extreme, in that it has a large number of hashtags (specifically, 11), some of them representing lexical variations of the same base word. 

These examples show not only that people use a variety of ways to hashtag tweets or that they may not understand or know how to hashtag tweets, but also that the user-provided hashtags tend to be either too general or too specific. 
Moreover, these examples are not exceptions, but rather they are representatives for a disaster-related tweet dataset. An analysis of a large corpus of tweets that was used in this work revealed that  most of the hashtags in a disaster-related tweet corpus simply represent disaster names and locations, and that approximately half of the tweets do not have any hashtags at all. 
Thus, filtering based on user-provided hashtags is not helpful for disaster response or people on the ground to quickly find relevant information. Similarly, filtering tweets during disasters based on keyword match is not expected to work well since keywords can be ambiguous and can lead to noisy results, e.g., a search for the keyword ``Harvey'' will retrieve tweets about the hurricane, but also about people whose name is ``Harvey.''  

To address these limitations of tweet retrieval based on user-provided hashtags or keywords, we envision a system that learns to identify relevant and topically informative hashtags and extract them directly from the content of the disaster tweets, capturing three main elements in a tweet: 1) disaster name; 2) location; and 3) situational awareness information.
%
Examples of relevant and topically informative hashtags that represent these three elements for the tweets in Table \ref{examples} are provided with each tweet and are shown in a light red box in the table (these hashtags are extracted directly from the tweets' content). 
Precisely, the first two tweets may be retrieved based on a search for \#Houston and \#needhelp. The third tweet can be retrieved with a search for \#HurricaneIrma, \#Orlando, \#powerlines.  The fourth tweet can be retrieved with a search for \#HurricaneFlorence, \#Wilmington and \#damage (or \#powerlines).  
Finally, the fifth  tweet can be retrieved when searching for \#WoolseyFire, \#ThousandOaks or \#Camarillo, or \#evacuated, and can be used to  find information about evacuated people. 
Thus, together, the above three elements can be used to filter tweets of potential interest to an emergency organization, which is responding to the disaster in question, or can be used to recommend hashtags in real time as the user types. 
Although there are previous works that focus on hashtag recommendation \cite{gong2016hashtag,li2016tweet,zhang2017hashtag,LI2019356} and topical keyphrase extraction \cite{marujo2015automatic,zhang2016keyphrase} for the general Twitter, research on identifying and extracting hashtags from the disaster Twitter data is limited. A notable exception is the work by \citeauthor{imran2013extracting} \shortcite{imran2013extracting}, where the authors extracted short information nuggets representing ``what'', ``where'', ``when'', etc., for a very small number of tweets classified in specific situational awareness categories. 

One potential reason that hindered progress on automatic hashtag identification (or extraction)\footnote{We use interchangeably hashtag identification and hashtag extraction in this paper.} from disaster-related tweets is the lack of large publicly available social media datasets annotated with relevant and topically informative hashtags. 
To fill in this gap, we constructed a large and unique dataset of disaster-related tweets annotated with hashtags to enable the development of deep learning techniques for automatic hashtag identification in order to further research in this critical area.

In doing so, we first collected tweets related to multiple disasters and disaster types (e.g., hurricane, flooding) and then manually crafted a lexicon, which was used together with the hashtags from the tweets, whenever available, to annotate a large dataset. Using this dataset, we further investigated a powerful deep learning model, initially proposed by \cite{zhang2016keyphrase} for keyphrase extraction from general tweets, and evaluate its performance capability for hashtag extraction from disaster-related tweets. This model, a joint-layer Long-Short Term Memory network trained using Multi-Task Learning (LSTM-MTL) and its variants that capture specifics of informal writing can be regarded as strong baselines on this dataset. 
Specifically, we make the following contributions:
\setlength{\parskip}{0em}
\begin{enumerate}
\item We present a hashtag annotated dataset of more than $67,288$ tweets related to disasters of various types (e.g., hurricanes, floodings, and earthquakes) and validate the hashtag annotations using human judgements. 
\item We develop an LSTM-MTL model and  variants that incorporate informal writing styles in order to exploit our new dataset for automatic hashtag extraction. 
The dataset, code, and other resources from this work are made available on Github.  
\item We conduct a thorough empirical evaluation of the LSTM-MTL model and its variants and show improvements of these variants over strong baselines. 
\end{enumerate}


\section{Related Work}
\label{related}


\citeauthor{mizuno2016wisdom} (\citeyear{mizuno2016wisdom}) described two systems that can be used to analyze and summarize information posted on social media during disasters. The first system, called  DISAANA,  can answer questions (e.g., ``What is in short supply in Kumamoto?'') and list problem reports  identified using Twitter data (e.g., ``people were burried alive''). The second system, called D-SUMM, can be used to summarize similar problem reports into a broader category. These systems work directly on Twitter, and could benefit from informative extractive hashtags in order to reduce the number of tweets they can be used on, and consequently, improve their speed. 

A variety of hashtag recommendation approaches have been proposed for general tweets. For example, \citeauthor{li2016tweet} (\citeyear{li2016tweet}) used a multi-class classification approach (i.e., each candidate hashtag represents a class), which combines the skip-gram model for finding word embeddings \cite{mikolov2013distributed}, with a convolutional neural network for learning sentence vectors, and a long short-term memory (LSTM) network \cite{Hochreiter:1997} for combining sentence vectors into a tweet vector. The tweet vectors were provided as input to a softmax layer, which was used to identify hashtags associated with the tweet, among a set of candidate hashtags. The approach was tested on a general tweet dataset, using a set of 20 popular hashtags. 
Experimental results showed that the proposed approach performed better than baselines that used TF-IDF or other types of recurrent neural networks. 

\citeauthor{gong2016hashtag} (\citeyear{gong2016hashtag}) also formulated the problem as a multi-class classification task, and proposed an approach based on convolutional neural networks, 
seen as a global channel, combined with the attention mechanism, 
seen as a local channel, to recommend hashtags.  
The attention mechanism was used to identify tweet words that trigger hashtags. The model was tested on a general tweet dataset 
with user-provided hashtags as gold-standard, and gave significant improvements over several baselines that did not use deep learning models. 

\citeauthor{li2016hashtag} (\citeyear{li2016hashtag}) proposed an LSTM-based approach that uses the attention mechanism to incorporate the topic information \cite{blei2003latent} of the tweet into the model. Implicitly, the model finds associations between local hidden representations of the words and the global topic information of the tweet, and uses these associations to generate a representation that leads to useful topical hashtags when passed through 
a softmax layer. 
Most recently, \citeauthor{LI2019356} (\citeyear{LI2019356}) extended their previous approach to include a co-attention mechanism that models content and topic information simultaneously. The extended approach was inspired from another hashtag recommendation approach \cite{zhang2017hashtag} that used the co-attention mechanism to combine textual and visual information available in many tweets. 
More specifically, the tweet content was modeled using a bidirectional LSTM (Bi-LSTM) sequence encoder, while the tweet topic was modeled using the approach proposed in \cite{Zhao:2011}. Using a co-attention mechanism, a new content/topic representation is learned for each tweet. As with the other hashtag recommendation approaches, this approach was evaluated on general tweets. 
Experimental results showed significant improvements over several baselines, including the previous model in \cite{li2016hashtag}.

One common theme to the hashtag recommendation approaches reviewed above is that they formulate the problem as a multi-class classification task, where the hashtags are a priori established, and a softmax layer transforms the hidden tweet representation into a probability distribution over the hashtags.  Usually, there is a small number of candidate hashtags, e.g., 20 as in \cite{li2016tweet}. 

While these approaches allow recommendation of hashtags that do not appear in the tweet, the requirement of pre-selecting a fixed number of hashtag candidates makes the classification-based approaches impractical for our purpose. In a time of disaster, it is expected that there could a new set of emerging disaster-related entities that we may want to use as hashtags. In critical times, we cannot afford to collect new sets of candidate hashtag classes and re-train the classification models. Therefore, instead of a classification-based approach, we take an extractive approach to this task. Precisely, we explore models for extracting important terms (which may serve as good hashtags) present in the tweets. As such, our task more closely aligns with keyphrase extraction. 

Hashtags in tweets are closely related to keyphrases. For example, \citeauthor{zhang2016keyphrase}\shortcite{zhang2016keyphrase} used hashtags as gold keyphrases for keyphrase extraction from Twitter. Therefore, we treat the task of hashtag identification as similar to the task of keyphrase extraction. Thus, close to our research is also the work on keyphrase extraction from Twitter \cite{marujo2015automatic,zhang2016keyphrase,zhao2011topical,bellaachia2012ne}. For example, \cite{marujo2015automatic} formulated the problem as binary classification and
showed that word embeddings in a system such as MAUI \cite{medelyan2009human} perform better than the TF-IDF \cite{sparck1972statistical} for keyphrase extraction on general tweets. 
\citeauthor{zhang2016keyphrase} \shortcite{zhang2016keyphrase} formulated the  problem as a sequence labeling task which allows the extraction of keyphrases of arbitrary lengths, without being constrained by some fixed number of classes. \citeauthor{zhang-etal-2018-encoding} \shortcite{zhang-etal-2018-encoding} extends the work of \citeauthor{zhang2016keyphrase} \shortcite{zhang2016keyphrase} by encoding conversational context. 



We chose to focus on a sequence labeling model for identifying hashtag, and specifically a model based on the Joint-Layer-RNN proposed by \citeauthor{zhang2016keyphrase} \shortcite{zhang2016keyphrase} since it achieved state-of-the-art performance on general tweets. This approach enables the model to extract new hashtags that may not have been seen before in the training data. This is particularly important in the context of disaster tweet annotation as new disasters with specific new names, locations, requests, and needs happen all the time. 

\begin{table}[h!]
\scriptsize
\centering
\def\arraystretch{1.3}
\begin{tabular}{  l | r } 
\toprule
\textbf{Disaster} & \textbf{Size}\\
\hline
Joplin Tornado \cite{imran2013extracting} & $2280$\\
\hline
Sandy Hurricane \cite{imran2013practical} & $514$ \\
\hline
Mexico Earthquake \cite{crisismmd2018icwsm} & $904$\\
Harvey Hurricane \cite{crisismmd2018icwsm} & $2952$\\
Irma Hurricane \cite{crisismmd2018icwsm} & $3050$\\
\hline
Nepal Earthquake \cite{firojACL2018embaddings} & $4890$\\
Queensland Floods \cite{firojACL2018embaddings}  & $2941$\\
\hline
Colorado Wildfires \cite{olteanu2015expect} & $623$\\
Costa Rica Earthquake \cite{olteanu2015expect} & $191$\\
Guatemala Earthquake \cite{olteanu2015expect} & $133$\\
Venezuela Refinery \cite{olteanu2015expect} & $36$\\
Alberta Floods \cite{olteanu2015expect} & $641$\\
Australia Bushfire \cite{olteanu2015expect} & $631$\\
Bohol Earthquake \cite{olteanu2015expect} & $334$\\
Boston Bombings \cite{olteanu2015expect} & $391$\\
Colorado Floods \cite{olteanu2015expect} & $696$\\
LA Airport Shootings \cite{olteanu2015expect} & $603$\\
Manila Floods \cite{olteanu2015expect} & $434$\\
Sardinia Floods \cite{olteanu2015expect} & $66$\\
Singapore Haze \cite{olteanu2015expect} & $281$\\
Typhoon Yolanda \cite{olteanu2015expect} & $619$\\
West Texas Explosion \cite{olteanu2015expect} & $422$\\
Typhoon Pablo \cite{olteanu2015expect} & $507$\\
Philippines Floods \cite{olteanu2015expect} & $540$\\
\hline
Pakistan Earthquake \cite{imran2016lrec} & $1262$\\
California Earthquake \cite{imran2016lrec} & $1453$\\
Chile Earthquake \cite{imran2016lrec} & $1047$\\
Ebola Virus \cite{imran2016lrec} & $1468$\\
Hurricane Odile \cite{imran2016lrec} & $992$\\
India Floods \cite{imran2016lrec} & $1183$\\
M.E.R.S \cite{imran2016lrec} & $1234$\\
Pakistan Floods \cite{imran2016lrec} & $1524$\\
Typhoon Hagupit \cite{imran2016lrec} & $1453$\\
Cyclone Pam \cite{imran2016lrec} & $1328$\\
Nepal Earthquake \cite{imran2016lrec} & $28$\\
\hline
Chiapas Earthquake (ours) & $2549$\\
Mexico Earthquake (ours) & $3742$\\
Harvey Hurricane (ours) & $4349$\\
Irma Hurricane (ours) & $4246$\\
Maria Hurricane (ours) & $12200$\\
California Fire (ours) & $2551$\\
\hline
All Disasters Tweets (in total) & $67,288$\\
\bottomrule
\end{tabular}
\caption{ Number of tweets extracted for different disasters from different sources.}
\label{table: 99}
\end{table}

\section{Hashtag Annotated Dataset}
\label{datasets} 

Previous approaches for tweet 
keyphrase or hashtag extraction have been used with general tweets, where the user-provided hashtags are considered to be the gold-standard. There is no large corpus for disaster tweet hashtag extraction. Furthermore, as explained above, the user-provided hashtags for disaster-related tweets do not always contain useful or sufficient information in terms of situational awareness and disaster response. Therefore, the strongest contribution of our work is to construct a benchmark dataset for disaster tweet hashtag extraction and to provide several models that can be used as strong baselines for this dataset.

\subsection{Collection of Disaster Tweets}
To construct a large ground-truth dataset for hashtag extraction, we first collected publicly available datasets that contain tweets related to a big variety of disasters of different types
\cite{imran2013practical,imran2013extracting,OlteanuCDV14,olteanu2015expect,imran2016lrec,firojACL2018embaddings,crisismmd2018icwsm}. All these existing datasets were annotated using crowdsourcing platforms 
or volunteers, with respect to relevance to a disaster and/or informativeness. We removed tweets annotated as irrelevant, as uninformative, or as related to emotional support and sympathy. 

\begin{figure}[t]
  \centering
  \fbox{\includegraphics[width=22em]{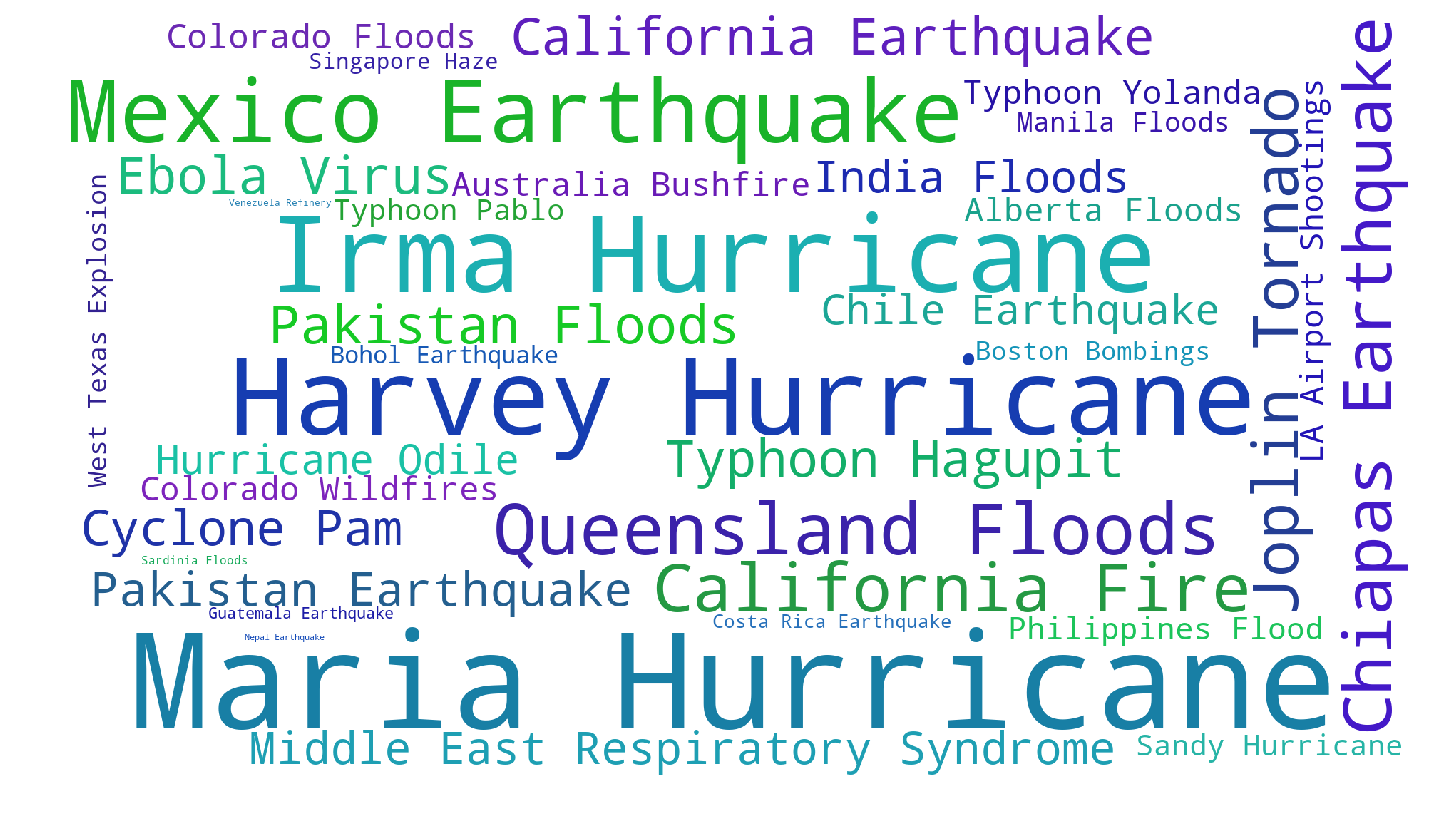}}
  \caption{Wordcloud of disasters in our dataset. }
  \label{fig:2}
\end{figure}

In addition to the publicly available disaster datasets, we also used a collection of tweets that we crawled, using the Twitter streaming API, during the following disasters: Hurricane Harvey, Hurricane Irma, Hurricane Maria, Mexico Earthquake, Chiapas Earthquake, and California Fire (wildfire).  
This collection was filtered for disaster relevance using a Na\"ive Bayes classifier trained on CrisisLexT6 \cite{OlteanuCDV14}. 
Finally, we  removed duplicates and non-English tweets from the combined dataset that consists of the publicly available datasets and our own collection of tweets. 
The resulting combined dataset comprises a large variety of disasters of different  types (37 disasters in total), 
as illustrated in Figure \ref{fig:2}. The size of each disaster name in the figure is proportional to the number of tweets from that disaster included in our new dataset. 
The overall details of our dataset including all the disasters present in it along with their corresponding source and size are provided in Table \ref{table: 99}.

\subsection{Disaster Lexicon}
Following the disaster-relevant data collection, we proceeded with a disaster lexicon construction, as follows: we first sampled approximately $150$ tweets from each of the $37$ disasters in our collection; then, we manually identified terms that were relevant and topically informative for the respective disaster and added them to the lexicon. The types of terms and phrases that we identified as relevant and informative are: location terms, requests or help-related terms (need food, send help, etc.), situation awareness terms (‘building collapse’, ‘100 dead’, ‘stranded’, ‘trapped’, etc.), disaster names (‘Hurricane’, ‘Hurricane Maria’), names of important persons, organizations, or nouns related to important vocations (related to ‘government’, ‘president’, etc.), terms related to disaster warnings and updates, and other terms associated  with infrastructure damage, property loss, affected individuals, injured or missing persons, fatality rates, etc. 

\setlength{\floatsep}{1em}
\setlength{\textfloatsep}{1em}

We only stored unigram and bigram phrases in the lexicon. In total, we extracted $2,140$ lexicon phrases from our sample tweets ($\approx 5,500$ tweets).
Finally, we included some phrases from the CrisisLex lexicon \cite{OlteanuCDV14}, which 
were not already in our lexicon. Our final lexicon contains $2,430$ unique phrases. 

\subsection{Dataset Annotation}
We used the manually constructed lexicon to automatically annotate {\it hashtags} in the tweets, by matching the lemmatized version of a tweet phrase with a lemmatized phrase in the disaster lexicon. We chose to use the lemmatized version of a word instead of its stemmed version to avoid ambiguous words resulting from chopping off the end of words.  
While our lexicon only contain bigrams and unigrams, we chain together overlapping bigram matches in a sequence to create larger keyphrases. Consider the example where we have a subsequence: ``hurricane maria recovery efforts", and we have the following bigram phrases from the lexicon: ``hurricane maria", ``maria recovery", and ``recovery efforts". Thus, there are overlapping matches between the given subsequence and the bigram lexicon phrases. In this case, we can chain the bigram matches together, combining them into a single annotated keyphrase: ``hurricane maria recovery efforts". 
We found that most of the time, phrases from the lexicon appear in the tweets within a similar context (due to the fact that we pre-filtered most tweets that do not occur in a  disaster context). 
This fact mitigates the risk of annotating phrases in an unsuitable context even though our annotation approach does not take context into account. 

In addition to hashtags annotated as gold-standard based on the manually constructed lexicon, we also used user-provided hashtags as gold-standard (when available), as they generally capture disaster names and locations, as mentioned before. However, we removed the \# sign, and segmented all hashtagged phrases into the constituent words, before annotating them, to ensure that the model learns to distinguish hashtag-like words without relying on the \# sign. We also removed user mentions and urls from the tweets. 

\setlength{\textfloatsep}{1em}
\setlength{\floatsep}{1em}

\subsection{Benchmark Dataset}
To enable progress on hashtag identification in disaster tweets and facilitate models' comparisons, we created a benchmark dataset by splitting our dataset into training, validation and test subsets. The test subset consists of: (1) three disasters that are not represented in the training data, specifically, Maria Hurricane, Philippines Flood and California Fire, and (2) $7\%$ of the data (removed from the training set) from the disasters represented in the training set. The performance of the models is evaluated on each of these four subsets separately. The validation subset consists of the whole Typhoon Pablo data, together with $15\%$ of the data (removed from the training set) from the disasters occurring in the training set.  

\subsection{Quality Assessment}
To assess the quality of our semi-automated lexicon annotation, we had human annotators manually inspect the lexicon-based annotations for a sample of tweets, and make an assessment about the annotation of a tweet as appropriate or not appropriate. For this task, we sampled $500$ 
tweets from our dataset (training set), and uploaded them to Amazon Mechanical Turk. 
The task was to decide whether a given predicted keyphrase is appropriate (as hashtag) for a given tweet or not. The specific options were ``Appropriate'', ``Not Appropriate'', and ``Unsure''. Each tweet was assessed by three annotators. Approximately $89\%$ of the keyphrases had a majority vote for ``Appropriate'' and only about $3\%$ of the sampled data were consensually voted as ``Not Appropriate'' by all the three annotators.

\setlength{\floatsep}{1em}
\setlength{\textfloatsep}{1em}
\begin{figure}[t]
  \centering
  \includegraphics[width=15em]{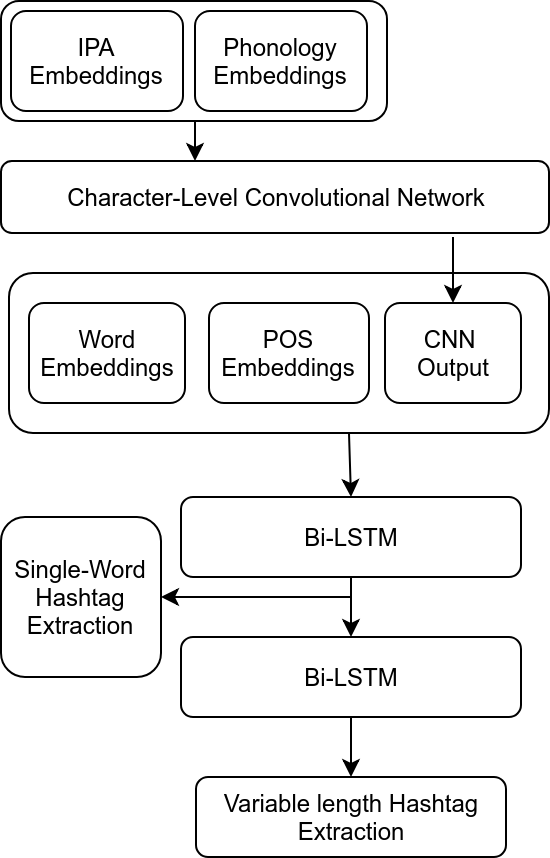}
  \caption{Diagram of the LSTM-MTL model with word embeddings, POS-embeddings, and concatenated IPA and phonological features.}
  \label{fig:1}
\end{figure}

\begin{table*}[h]
\small
\centering
\def\arraystretch{1.2}
\begin{tabular}{ l | c  c  c | c  c  l } 
\toprule
\textbf{Model} & \textbf{Pr} & \textbf{Re} & $\mathbf{F_1}$ & \textbf{Pr} & \textbf{Re} & $\mathbf{F_1}$ \\
\midrule
& \multicolumn{3}{c|}{Maria Hurricane} & \multicolumn{3}{c}{California Fire}\\
\midrule
LSTM & $89.65\%$ & $83.49\%$ & $86.46\%$ & ${\underline {91.21\%}}$ & $85.29\%$ & ${\underline {88.15\%}}$\\
2-layer LSTM & ${\underline {89.81\%}}$ & $83.08\%$ & $86.31\%$ & $91.09\%$ & $84.79\%$ & $87.83\%$\\
LSTM-MTL & ${\underline {89.87\%}}$ & ${\underline {83.93\%}}$ & ${\underline {86.80\%}}$ & $90.96\%$ & ${\underline {85.40\%}}$ & $88.09\%$\\
\hline
LSTM-MTL+ELMo & $90.48\%$ & ${\underline {\mathbf{85.80\%}}}$ & $88.08\%$ & ${\underline{\mathbf{92.65\%}}}$ & ${\underline {\mathbf{89.19\%}}}$ &  ${\underline {\mathbf{90.89\%}}}$\\ 
LSTM-MTL+IPA,POS & $89.19\%$ & $84.92\%$ & $87.00\%$ & $91.00\%$ & $86.76\%$ & $88.83\%$\\
LSTM-MTL\newline+ELMo,IPA,POS & ${\underline {\mathbf{90.97\%}}}$ & $85.59\%$ & ${\underline {\mathbf{88.20\%}}}$ & $91.88\%$ & $87.72\%$ & $89.75\%$\\

\midrule
& \multicolumn{3}{c|}{Philippines Flood} & \multicolumn{3}{c}{Multiple disasters}\\
\midrule
LSTM & ${\underline {85.99\%}}$ & $83.39\%$ & $84.67\%$ & ${\underline {92.89\%}}$ & $88.95\%$ & $90.88\%$\\
2-layer LSTM & $85.71\%$ & $83.59\%$ & ${\underline {84.64\%}}$ & $92.39\%$ & $88.08\%$ & $90.18\%$\\
LSTM-MTL & $85.29\%$ & ${\underline {83.94\%}}$ & $84.61\%$ & $92.88\%$ & ${\underline {89.27\%}}$ & ${\underline {91.04\%}}$\\
\hline
LSTM-MTL+ELMo &  $87.30\%$ & $85.52\%$ & $86.40\%$ & $93.67\%$ & $90.55\%$ & $92.08\%$ \\ 
LSTM-MTL+IPA,POS &  $87.26\%$ & $84.58\%$ & $85.90\%$ & $93.38\%$ & $89.74\%$ & $91.52\%$\\
LSTM-MTL\newline+ELMo,IPA,POS & ${\underline {\bf 87.88\%}}$ & ${\underline {\bf 85.92\%}}$ & ${\underline {\bf86.89\%}}$ & ${\underline {\bf 93.83\%}}$ & ${\underline {\bf 90.66\%}}$ & ${\underline {\bf 92.22\%}}$ \\
\bottomrule
\end{tabular}
\caption{Precision, Recall, and F1 scores on four test datasets}
\label{table:5}
\end{table*}
\setlength{\parskip}{0.5em}

\setlength{\floatsep}{2em}
\setlength{\textfloatsep}{2em}
\begin{table*}[h]
\small
\centering
\def\arraystretch{1.2}
\begin{tabular}{  |p{50em}| } 
\toprule
\rule{0pt}{3.6ex}\colorbox{blue_}{\strut Flood} in the \rule{0pt}{3.6ex}\colorbox{orange_}{\strut UST }\colorbox{blue_}{\strut Hospital} is now on the 2nd floor. \colorbox{blue_}{\strut No food} for the patients \& staff. \colorbox{blue_}{\strut Pls. help!} \colorbox{blue_}{\strut \#rescuePH} @norescu\rule[-2.6ex]{0pt}{0pt}\\
\hline
\rule{0pt}{3.6ex}was so happy to see these two \colorbox{blue_}{\strut babies} \colorbox{blue_}{\strut pulled out} of \colorbox{blue_}{\strut collapsed building} \colorbox{blue_}{\strut alive} heartbreaking bawling my eyes out \rule[-2.6ex]{0pt}{0pt}\\
\hline
\rule{0pt}{3.6ex}\colorbox{blue_}{\strut Hurricane Maria} came ashore in \colorbox{blue_}{\strut Puerto Rico} this morning as a \colorbox{blue_}{\strut category 4} \colorbox{blue_}{\strut hurricane} with winds of \colorbox{blue_}{\strut 55 mph}\rule[-2.6ex]{0pt}{0pt}\\
\hline
\rule{0pt}{3.6ex}\colorbox{blue_}{\strut 14 people killed} and several \colorbox{blue_}{\strut missing} after \colorbox{blue_}{\strut Cyclone Cleopatra} hit \colorbox{blue_}{\strut Italian island} of \colorbox{blue_}{\strut Sardinia}, officials say \rule[-2.6ex]{0pt}{0pt}\\
\hline
\rule{0pt}{3.6ex}\colorbox{blue_}{\strut pls help: People} in Hermosa, \colorbox{red_}{\strut Bataan} r in roofs now,there's no \colorbox{blue_}{\strut rescuers}\colorbox{orange_}{\strut }\colorbox{blue_}{\strut helping} as of now \colorbox{blue_}{\strut \#rescuePH}\rule[-2.6ex]{0pt}{0pt}\\
\hline
\rule{0pt}{3.6ex}Maybe this will \colorbox{blue_}{\strut help}. Please \colorbox{blue_}{\strut donate} \colorbox{red_}{\strut Hispanic Federation} and direct \colorbox{blue_}{\strut relief} online highly rated by \colorbox{blue_}{\strut charity} naviga\rule[-2.6ex]{0pt}{0pt}\\
\hline
\rule{0pt}{3.6ex}\colorbox{blue_}{\strut \#rescuePH} \colorbox{blue_}{\strut Pregnant} mom and small \colorbox{blue_}{\strut kids}\colorbox{red_}{\strut }\colorbox{blue_}{\strut trapped} in 5 feet \colorbox{blue_}{\strut flooded} house 
\colorbox{orange_}{\strut valenzuela city}.\rule[-2.6ex]{0pt}{0pt}\\
\bottomrule
\end{tabular}
\caption{Examples of tweets from our test dataset. The agreement between predicted hashtags and annotated gold-standard hashtags is marked with blue. Gold-standard hashtags that are not in the set of predicted hashtags are marked with yellow, and predicted hashtags that are not annotated as gold-standard are marked with red. The predictions are made with LSTM-MTL{\footnotesize+ELMo, POS, IPA} model. (Personal information and urls were removed.)}
\label{table:4}
\end{table*}

\section{LSTM and Variants}
\label{models}

For the core modeling of our dataset, we use 
the Joint-layer Recurrent Neural Network (RNN) model proposed by \citeauthor{zhang2016keyphrase} \shortcite{zhang2016keyphrase} and investigate several of its variants that capture specifics of informal writing in social media. We chose this model because it achieves state-of-the-art performance on general Tweet dataset (with hashtags treated as ground truth). It significantly outperforms other models such as Marujo's \cite{marujo2015automatic} variant of MAUI \cite{medelyan2009human}. Furthermore, 
traditional keyphrase extraction models such as TF-IDF \cite{Salton:1986:IMI:576628}, TextRank \cite{mihalcea2004textrank}, or KEA \cite{Witten:1999:KPA:313238.313437} that rely on statistical features like word co-occurences and word counts of terms within a document, are not expected to work well on Twitter data since tweets consist of very short text.
Hence, in a tweet, most candidate keyphrases will usually occur only once and a word can only co-occur with very few other different words. Indeed, we find them to perform considerably poorly compared to Joint-layer RNN based models on Twitter data in the work of \citeauthor{zhang-etal-2018-encoding} \shortcite{zhang-etal-2018-encoding}. We also compare the Joint-layer RNN with other RNN models (a single layered BiLSTM, and a two layered BiLSTM). We describe Joint-layer RNN model and its variants in what follows. A diagram of the most complete variant is shown in Figure \ref{fig:1}. \setlength{\parskip}{1em}

\noindent 
\textbf{LSTM-MTL:} The Joint-layer RNN is a Bi-LSTM model \cite{Hochreiter:1997,graves2013hybrid}, trained using Multi-Task Learning (MTL), 
which stacks two Bi-LSTM layers and jointly trains them on two related tasks. The first Bi-LSTM is trained on the task of identifying single words that are suited to be part of a hashtag (a lower level auxiliary task). The second Bi-LSTM is trained to label hashtag candidate phrases of arbitrary length (the main task, treated as a sequence labeling problem). 
Similar MLT approaches have been used in other contexts \cite{sogaard2016deep,liu2016recurrent}. \setlength{\parskip}{0em}
For this model, we used GloVe embeddings \cite{pennington2014glove} pre-trained on Twitter. The embeddings were loaded using Flair 
 \cite{akbik2018coling}.  Following  \citeauthor{zhang2016keyphrase} \shortcite{zhang2016keyphrase}, we  
represented each word in a sequence as a concatenation of three words, specifically, the  current word and its immediate neighbors. 

\setlength{\parskip}{0.5em}
\noindent {\bf LSTM-MTL{\footnotesize+ELMo}:} In this variant of the Joint-layer RNN, we concatenate the GloVe embeddings with contextualized ELMo word embeddings  \cite{peters2018deep}. 
However, to keep the number of parameters smaller, we did not use the three-word window representation as in the original model, as ELMo is already encoding the context. 

\setlength{\parskip}{0.5em}
\noindent {\bf LSTM-MTL{\footnotesize+IPA,POS}:} This model variant of the Joint-layer RNN is aimed at better handling noise in the data. Hence, we incorporate information about the informal writing, inspired from the methods proposed by \citeauthor{aguilar2018modeling} \shortcite{aguilar2018modeling}. They noted how Twitter users often tend to spell words based on their pronunciations, which means it is possible to make more normalized representations of the words by using their phonetics or corresponding IPA (International Phonetic Alphabet) letters, alongside with their phonological features. Following their work, we used Epitran
\cite{Mortensen-et-al:2018} to convert graphemes to phonemes (represented using IPA), and Panphon \cite{Mortensen-et-al:2016} to convert each IPA phoneme into a vector representing various phonological (articulatory) features associated with it. We also used part-of-speech (POS) taggers as provided by \citeauthor{owoputi2013improved} \shortcite{owoputi2013improved} to explicitly add POS-tag information to this model. 
\setlength{\parskip}{0em}


We used randomly initialized embeddings for each POS-tag and IPA symbol. We directly used the phonological vector representations created with Panphon as embeddings for phonological features. We concatenated the embeddings of IPA symbols with their corresponding phonological feature vectors. The result of the concatenation, a character level representation of the phonetics and phonological features for each word, was fed to a character-level CNN \cite{zhang2015character}, followed by a global max-pooling layer, to create word level representations. Unlike \citeauthor{aguilar2018modeling} (\citeyear{aguilar2018modeling}), we chose to use a CNN as opposed to a Bi-LSTM for creating the word level representations because the CNN is much faster. The output of the CNN was concatenated with the POS-tag embeddings and  pre-trained GloVe embeddings. The resultant representation was then fed to the stacked Bi-LSTM model. This model uses the three-word window representation on GloVe embeddings. 
\setlength{\parskip}{0.5em}

\noindent {\bf LSTM-MTL{\footnotesize+ELMo,IPA,POS}:} The last variant of the Joint-layer RNN is a combination of the above two models. It uses ELMo concatenated with GloVe embeddings, POS-tag embeddings, and CNN encoded word-level representations of phonetics and phonological features. Given that ELMo captures context, this model does not use the three-word window representation.  \setlength{\parskip}{0em}

\section{Experimental Setup and Results}
We describe our experimental setting and present  results on our four test datasets in this section. 
\label{experiments and results}

\subsection{Experimental Settings} 
\label{setting}

We used $100$ dimensional Twitter GloVe embeddings, $1024$ dimensional ELMo embeddings, $64$ dimensional POS-tag embeddings, and $22$ dimensional IPA embeddings. Phonological features were represented with a $22$ dimensional vector.  The embeddings were further fine-tuned during training. Each Bi-LSTM network had $300$ hidden units. For the CNN, we used $128$ filters and a kernel of size $3$. We used dropouts of $0.5$ on the input to the Bi-LSTM layers. For optimization, we used the nadam optimizer \cite{dozat2016incorporating} with a learning rate of $0.0015$. Hyper-parameters were either tuned on the validation data or selected based on values that gave good results in prior works. 

\subsection{Results} 
\label{Results}
The results of the experiments are shown in Table  \ref{table:5} for the four test datasets, specifically, Maria Hurricane, California Fire, Philippines Floods, and a dataset sampled from multiple disasters that are explicitly represented in the training data (however, with no overlap between train and test). The multiple disasters dataset is used to evaluate the ability of the models to generalize between similar training and test data, and can be seen as providing an upper bound for the performance of the models. We grouped the results according to the baselines (LSTM, 2-layer LSTM, and LSTM-MTL) and our explored variants of the LSTM-MTL that capture informal writing in social networks.
Underlined scores in the table are best within each group and bold scores are best overall. 
%
As can be seen from Table \ref{table:5}, our models can generalize well to unseen and underrepresented disasters like California Fire, Philippines Floods, and Maria Hurricane. 

Regarding the performance of the models, interestingly, we find in Table \ref{table:5} that the simpler LSTM models (LSTM and 2-layer LSTM) perform on par with the LSTM with multi-task setup (LSTM-MTL). 
However, the variants that explicitly incorporate specifics from the informal writing bring further improvements, with the LSTM-MTL{\footnotesize+ELMo,IPA,POS} model being the best overall (with the exception of California Fire dataset). Between the LSTM-MLT{\footnotesize+ELMO} and LSTM-MLT{\footnotesize+IPA,POS} variants, the LSTM-MLT{\footnotesize+ELMO} variant gives better results overall. We can also observe that the concatenation of the GloVe embeddings with the contextual ELMo embeddings capture better the context of the tweet  as compared to the LSTM-MTL model. 

\subsection{Error Analysis and Prediction Quality}
To gain insights into the hashtag predictions made by the best performing model, LSTM-MTL{\footnotesize+ELMo,IPA,POS}, Table \ref{table:4} shows the predictions on several tweets from our test data, by comparison with the gold-standard annotations. The agreement between model predictions and gold-standard is shown in blue. Predicted hashtags that are not in the gold-standard are marked with red, while gold-standard annotations that are not in the set of predicted hashtags are marked with yellow. As can be seen, the agreement between model predictions and gold-standard is very high. Interestingly our model can predict certain named entities such as `Hispanic Federation' and '`ataan,' which were beyond the lexicon, but it also misses some other named entities such as `valenzuela city', and `UST Hospital'. Overall we can see that the chosen model that capture informal writing is  able to make good hashtag predictions for new disasters.

The models described here can be integrated into systems that can help response organizations to have a real time map of a disaster - what is happening on the ground, which could display both the physical disaster and the spikes of intense activity in the proximity to the disaster. In time, such AI-based systems could have a strong social impact with great benefits to those affected by disasters, and could 
pinpoint the joy of having survived a falling tree, the horror of a bridge washing out, or the fear of looters
in action. Responders will be able to use such a system to provide real time alerts of the status of the disaster and of the affected population.

\section{Conclusion}
\label{concl}

In this paper, we introduced a new disaster-related dataset of tweets that were annotated with hashtags using a semi-automated lexicon-based approach.  
This is the first large-scale dataset constructed for identifying relevant and topically informative hashtags for disaster tweets. We believe that our dataset will foster research in this domain, will enable the design of deep learning models, and will help response organizations to make better use of social media data contributed by individuals affected by disasters, and will contribute to better decision-making during disasters when resources are limited. In addition to introducing a new dataset,  we built an LSTM-MTL model and explored its variants to capture informal writing in social media. The results show that taking informal writing into account improves the F1-score of LSTM-MTL by up to $2\%$. This opens up directions for future investigation for more explicitly capturing informal writing into the modeling. Also, incorporating domain knowledge into the models would be expected to improve performance further. In the near future, we wish to bring more attention to building a multilingual setup.
Hierarchical multiclass classification with fine and coarse grained labels enabled by our dataset 
would also be an interesting future direction. 

\subsubsection{Acknowledgments.}
We thank the National Science Foundation (NSF) and Amazon Web
Services for support from grants IIS-1741345 and IIS-1912887, which
supported the research and the computation in this study. We also
thank NSF for support from the grants IIS-1526542, IIS-1423337,
IIS-1652674, and CMMI-1541155. Any opinions, findings, and conclusions
expressed here are those of the authors and do not
necessarily reflect the views of NSF. We also thank our anonymous reviewers for their constructive feedback. 
\small{
\bibliography{aaai20}
\bibliographystyle{aaai}
}
\end{document}